\documentclass[11pt,leqno]{amsart}
\usepackage{amsthm,amsfonts,amssymb,amsmath,oldgerm}
\numberwithin{equation}{section}
\usepackage{graphicx}
\usepackage{fullpage}

\newcommand\br{\begin{remark}}
\newcommand\er{\end{remark}}
\newcommand\bp{\begin{pmatrix}}
\newcommand\ep{\end{pmatrix}}
\newcommand{\be}{\begin{equation}}
\newcommand{\ee}{\end{equation}}
\newcommand\ba{\begin{equation}\begin{aligned}}
\newcommand\ea{\end{aligned}\end{equation}}

\newcommand{\bap}{\begin{app}}
\newcommand{\eap}{\end{app}}
\newcommand{\begs}{\begin{exams}}
\newcommand{\eegs}{\end{exams}}
\newcommand{\beg}{\begin{example}}
\newcommand{\eeg}{\end{exaplem}}
\newcommand{\bpr}{\begin{proposition}}
\newcommand{\epr}{\end{proposition}}
\newcommand{\bt}{\begin{theorem}}
\newcommand{\et}{\end{theorem}}
\newcommand{\bc}{\begin{corollary}}
\newcommand{\ec}{\end{corollary}}
\newcommand{\bl}{\begin{lemma}}
\newcommand{\el}{\end{lemma}}
\newcommand{\bd}{\begin{definition}}
\newcommand{\ed}{\end{definition}}
\newcommand{\brs}{\begin{remarks}}
\newcommand{\ers}{\end{remarks}}

\newtheorem{theorem}{Theorem}[section]
\newtheorem{proposition}[theorem]{Proposition}
\newtheorem{corollary}[theorem]{Corollary}
\newtheorem{lemma}[theorem]{Lemma}

\theoremstyle{remark}
\newtheorem{remark}[theorem]{Remark}
\theoremstyle{definition}
\newtheorem{definition}[theorem]{Definition}

\newtheorem{example}[theorem]{Example}

\newcommand{\beq}{\begin{equation}}
\newcommand{\eeq}{\end{equation}}

\title{Note on the stability of viscous roll-waves}

\author{Blake Barker}
\address {Division of Applied Mathematics, Brown University, Providence, RI 02912 USA}
\email{Blake\underline{ }Barker@brown.edu}
\thanks{Research of B.B. was partially supported
under NSF grant no. DMS-1400872.}

\author{Mathew A. Johnson}
\address{Department of Mathematics, University of Kansas, Lawrence, KS 66045 USA}
\email{matjohn@ku.edu}
\thanks{Research of M.J. was partially supported under NSF grant no. DMS-1211183.
 }
 
\author{Pascal Noble}
\address{Institut de Math\'ematiques de Toulouse, INSA, Toulouse, France}
\email{pascal.noble@math.univ-toulouse.fr}

\author{L.Miguel Rodrigues}
\address{IRMAR--UMR CNRS 6625, Universit'e Rennes 1, F-35042 Rennes, France}
\email{luis-miguel.rodrigues@univ-rennes1.fr}
\thanks{Research of M.R. was partially supported by the ANR project
BoND ANR-13-BS01-0009-01.}

\author{Kevin Zumbrun}
\address{Department of Mathematics, Indiana University, Bloomington, IN 47405}
\email{kzumbrun@indiana.edu} 
\thanks{Research of K.Z. was partially supported
under NSF grant no. DMS-0300487.}

\begin{document}

\begin{abstract}
In this note, we announce a complete classification of
stability of periodic roll-wave solutions of the viscous shallow-water equations,
from their onset at Froude number $F\approx 2$ up to the infinite-Froude limit. 
For intermediate Froude numbers, we 
obtain numerically a particularly simple power-law relation between $F$ and the boundaries of the region of stable periods, that appears potentially
useful in hydraulic engineering applications. In the asymptotic regime $F\to 2$ (onset), we provide an analytic expression of the stability boundaries whereas in the limit $F\to\infty$, we show that 
roll-waves are always unstable. 
\end{abstract}
%
%

\date{\today}
\maketitle

\section{Introduction}\label{sec1}

In this note, we announce the classification in \cite{Barker2015,JNRZ2014,Rodrigues2015}
of spectral stability of roll-wave solutions of the ``viscous'' St. Venant equations for
inclined shallow-water flow, taking into account drag and viscosity. 
Written in nondimensional Eulerian form, the shallow water equations for a thin film down an incline are
\begin{equation}\label{swe1}
\displaystyle
\partial_t h+\partial_x(hu)=0,\qquad \partial_t(hu)+\partial_x\left(hu^2+\frac{h^2}{2\,F^2}\right)=h-|u|u+\nu\partial_x(h\partial_x u),
\end{equation} 
\noindent
where $F$ is the Froude number and $\nu=Re^{-1}$ is the inverse of the Reynolds number. 
Here $h(x,t)$ denotes the fluid height whereas $u(x,t)$ is the fluid velocity averaged with respect to height. The terms $h$ and $|u| u$ on the right hand side of the second equation model, respectively, gravitational force and turbulent friction along the bottom. Roll-waves are well-known hydrodynamic instabilities of (\ref{swe1}), arising in the region $F>2$ for which constant solutions, corresponding to parallel flow, are unstable. They are commonly found in man-made conduits such as aqueducts and spillways, and have been reproduced in laboratory flumes \cite{Brock1969}. 
However, up until now, there has been no complete rigorous stability analysis of viscous St. Venant roll-waves either at the linear (spectral) or nonlinear level.

Roll-waves may be modeled as periodic wave train solutions of (\ref{swe1}).
In \cite{JNRZ2014}, it was proved for a large class of viscous conservation laws and under suitable spectral assumptions that periodic wave trains are nonlinearly stable (in a spatially-modulated sense).   In 
\cite{Rodrigues2015,JNZ2011} this nonlinear analysis has been extended  to encompass all periodic
wave train solutions of the shallow water system (\ref{swe1}) that satisfy those spectral assumptions. The main issue then is the verification of such assumptions. Here, we provide a complete description of the set of stable roll-waves of (\ref{swe1}): for each Froude number $F>2$, we exhibit (either theoretically or numerically) the range of spatial periods where stable roll-waves are found. To our knowledge, this is the first complete result of stability in the case of shallow water equations. However, let us mention the study in \cite{Boudlal2002}: there, the authors studied the {\it modulational stability} of Dressler 
{\it inviscid} roll-waves. A set of modulation equations is derived by assuming that the parameters which encode the roll-waves slowly vary in time and space: 
lack of hyperbolicity of the modulation equations is expected to provide a sufficient criterion 
for spectral instability of roll-waves under special kinds of large scale perturbations.
%

In Section \ref{sec2}, we introduce the spectral problem and recall the spectral assumptions that have to be verified in order to obtain nonlinear stability of periodic waves. 
In Section \ref{sec3}, we consider the intermediate Froude number regime $2\leq F\leq 100$. We find a dramatic transition around $F\approx 2.3$ from the small-F description of stability to a  remarkably simple power-law description of surfaces 
bounding from above and below regions in parameter space corresponding to stable waves.  These surfaces 
eventually intersect, yielding instability for all sufficiently large $F$.  In Section \ref{sec4}, we focus on two asymptotic regimes: $F\to 2$ (onset) and $F\to\infty$. As $F\to 2$, the shallow water equations reduce to a generalized Kuramoto-Sivashinsky equation and we obtain asymptotic analytic formula for the stability boundaries. As $F\to\infty$, we exhibit a non-trivial regime and an asymptotic model which admits only unstable roll-waves, indicating the instability of roll-waves for
sufficiently large $F$.

\section{\label{sec2} Formulation of the spectral problem}

As the full nonlinear theory is given in Lagrangian coordinates of mass  \cite{JNZ2011,Rodrigues2015}, for the sake of consistency we rewrite the viscous shallow water system (\ref{swe1}) as
\begin{equation}\label{swe2}
\displaystyle
\partial_t \tau -\partial_x u=0,\quad \partial_t u+\partial_x\left(\frac{\tau^{-2}}{2\,F^2}\right)=1-\tau\,u^2+\nu\partial_x(\tau^{-2}\partial_x u),
\end{equation}
\noindent
where $\tau:=1/h$ and $x$ denotes now a Lagrangian marker rather than a physical location $\tilde x$, satisfying the relations $d\tilde x/dt = u(\tilde x, t)$ and  $dx/d\tilde x=\tau(\tilde x, t)$.  There is a one-to-one correspondence between periodic waves of the Lagrangian and Eulerian forms. It also holds for the spectral problem in its Floquet-by-Floquet description; see \cite{Benzoni2015}. Thus there is no loss of information in choosing to work with the Lagrangian form.
Now we introduce the spectral problem. Denote by 
$(\bar \tau, \bar u, \bar c)$ 
a particular periodic traveling (roll-wave) solution of (\ref{swe2}) of period $X$.
Linearizing (\ref{swe2}) about $(\bar\tau, \bar u)$ in the co-moving frame $(x-\bar c t,t)$ and 
seeking modes of the form $(\tau,u)(x,t)=e^{\lambda t}(\tau,u)(x)$, one obtains
\begin{equation}\label{spec1}
\begin{array}{ll}
\displaystyle
(u+\bar c\tau)'=\lambda\,\tau,\\
\displaystyle
\nu(\bar\tau^{-2}u')'=(\lambda+2\bar u\bar\tau) u-\left(\left(\frac{\bar\tau^{-3}}{F^2}-2\bar\tau^{-3}\bar u'\right)\tau'+\bar c u'\right)+\left(\bar u^2-\left(\frac{\bar\tau^{-3}}{F^2}-2\bar\tau^{-3}\bar u'\right)'\right)\tau,
\end{array}
\end{equation}
\noindent
where primes denote differentiation with respect to $x$.  Setting $v=(\tau,u)^T$, the spectral problem (\ref{spec1}) 
may be written as
$Lv=\lambda v$ where $L$ is a differential operator with {\it periodic coefficients}. By Floquet theory, one has
that $\lambda\in\sigma_{L^2(\mathbb{R})}(L)$ (the spectrum of $L$ acting on $L^2(\mathbb{R})$) if and only if there are  $\xi\in[-\pi/X,\pi/X)$ and $w\in L^2_{per}([0, X])$ 
(a function of period $X$) such that $L_{\xi}\,w=\lambda w$, where $L_\xi$ is the corresponding Bloch operator defined via
$
\left(L_\xi w\right)(x):=e^{-i\xi x}L\left[e^{i\xi\cdot}w(\cdot)\right](x).
$
Consequently, the spectrum may be decomposed into countably many curves $\lambda(\xi)$ of $L^2_{\rm per}([0,X])$-eigenvalues of the operators $L_\xi$. Roll-waves are proved to be nonlinearly stable under the following {\it diffusive spectral stability conditions}:
\begin{enumerate}
	\item (D1) $\sigma_{L^2(\mathbb{R})}(L)\subset\{\lambda\in\mathbb{C}\,|\,\Re(\lambda)<0\}\cup\{0\}$.
	\item (D2) There exists a $\theta>0$ such that for all $\xi\in[-\pi/X, \pi/X)$, $\sigma_{L^2_{per}([0, X])}(L_\xi)\subset\{\lambda\,|\,\Re(\lambda)\leq -\theta\xi^2\}$.
	\item (D3) $\lambda=0$ is an eigenvalue of $L_0$ with generalized eigenspace $\Sigma_0\subset L^2_{per}([0, X])$ of dimension $2$.
\end{enumerate}
\noindent
For a discussion of the significance of these conditions, see \cite{JNRZ2014}. In order to locate the spectrum, we introduce the Evans function $E_{SV}(\lambda,\xi)$. Write
(\ref{spec1}) as a first order differential system by setting $Z=(\tau, u, \bar\tau^{-2}u')^T$: $Z'=A(\cdot,\lambda)Z$.
Denoting the resolvent matrix $R(\cdot,\lambda)$ associated to this system, it follows 
that $\lambda\in\sigma_{L^2_{per}([0, X])}(L_\xi)$ 
if and only if $\lambda$ satisfies
$
\displaystyle
E_{SV}(\lambda,\xi):={\rm det}\left(R(X,\lambda)-e^{i\xi\, X}{\rm Id}_{\mathbb{R}^3}\right)=0. 
$

\section{\label{sec3} Numerical Study: Spectral stability for intermediate $F$}

\begin{figure}
	\begin{center}
		$
		\begin{array}{lcr}
		(a) \includegraphics[scale=0.25]{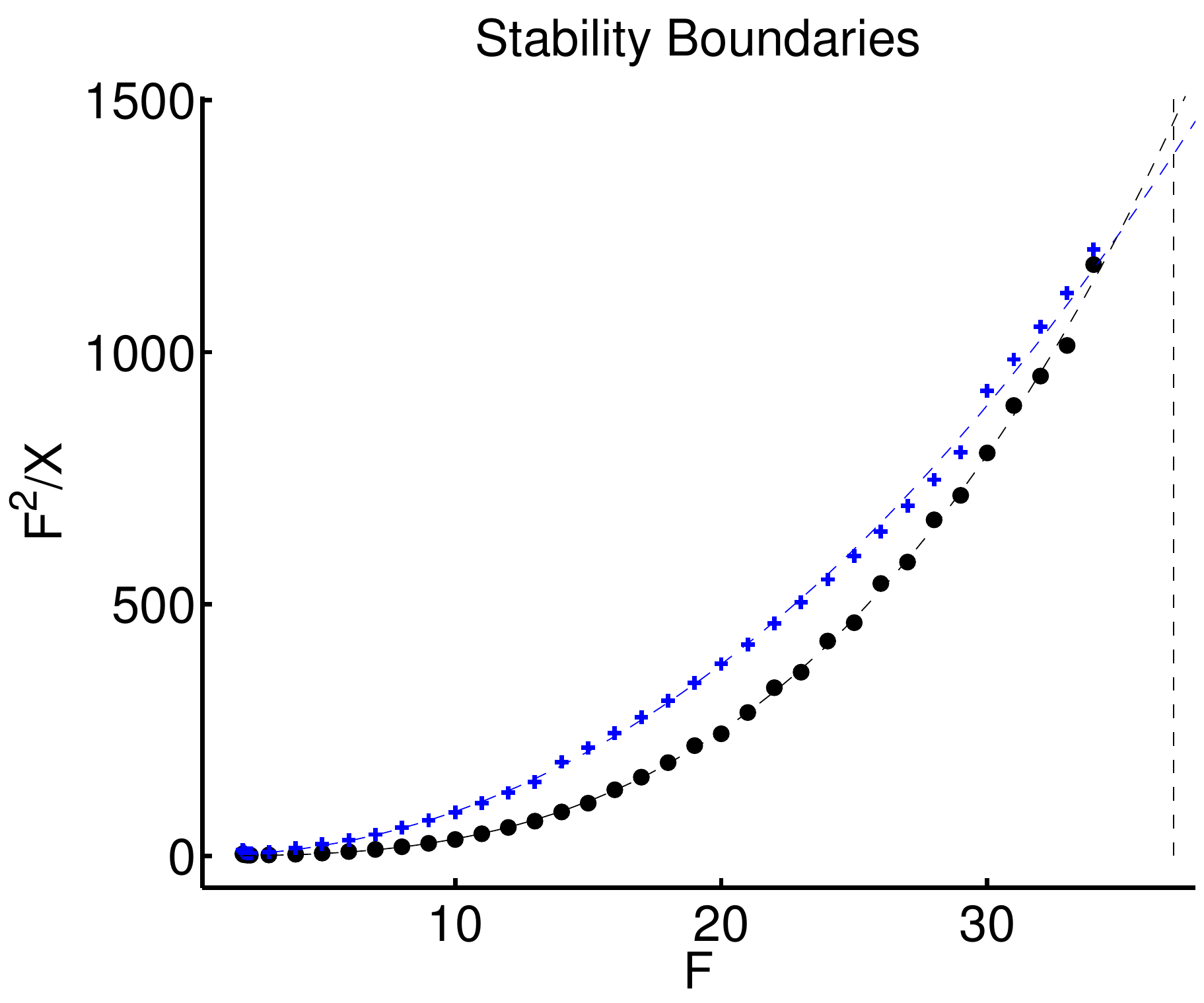}
		&(b) \includegraphics[scale=0.25]{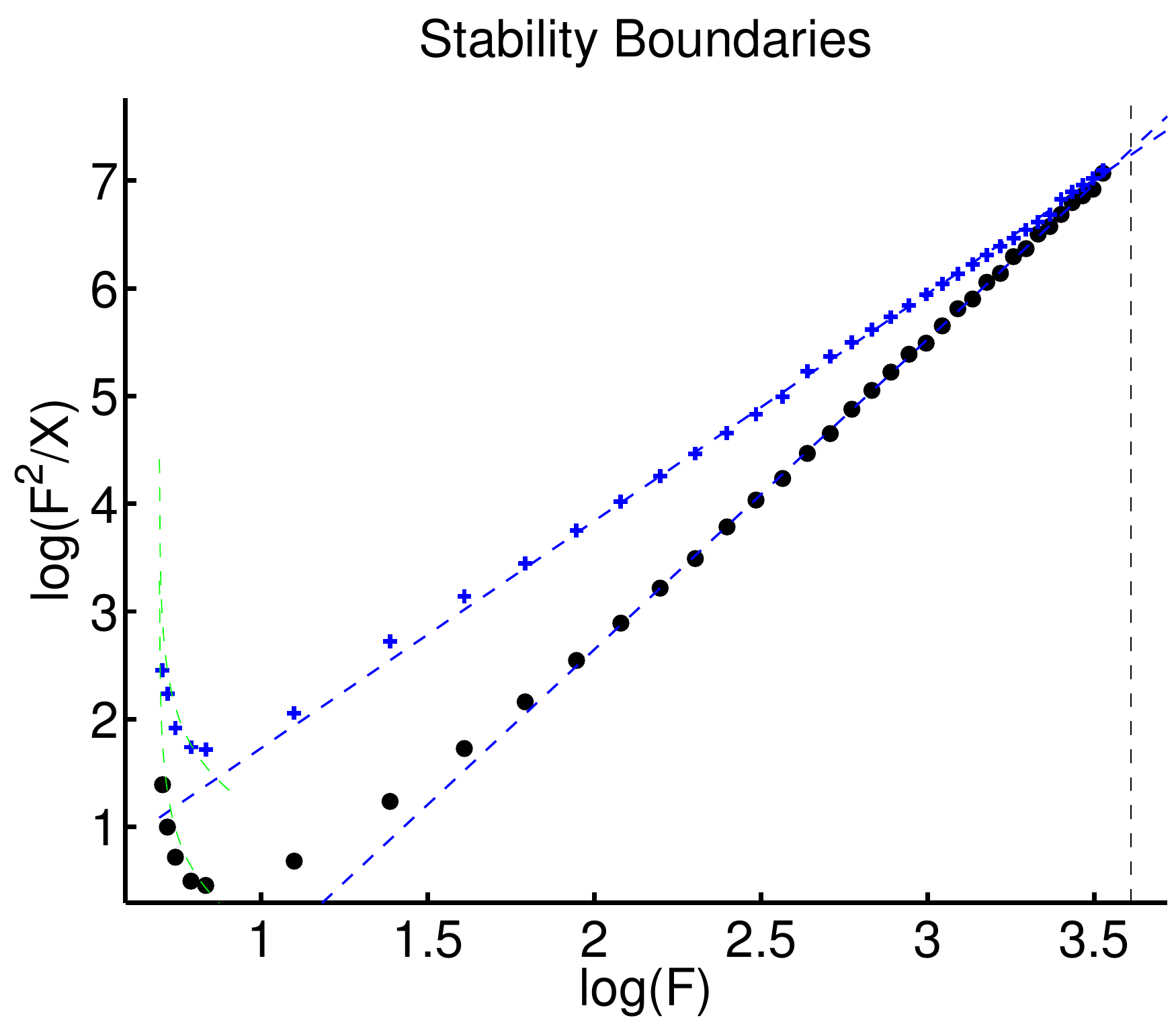}
		&(c) \includegraphics[scale=0.25]{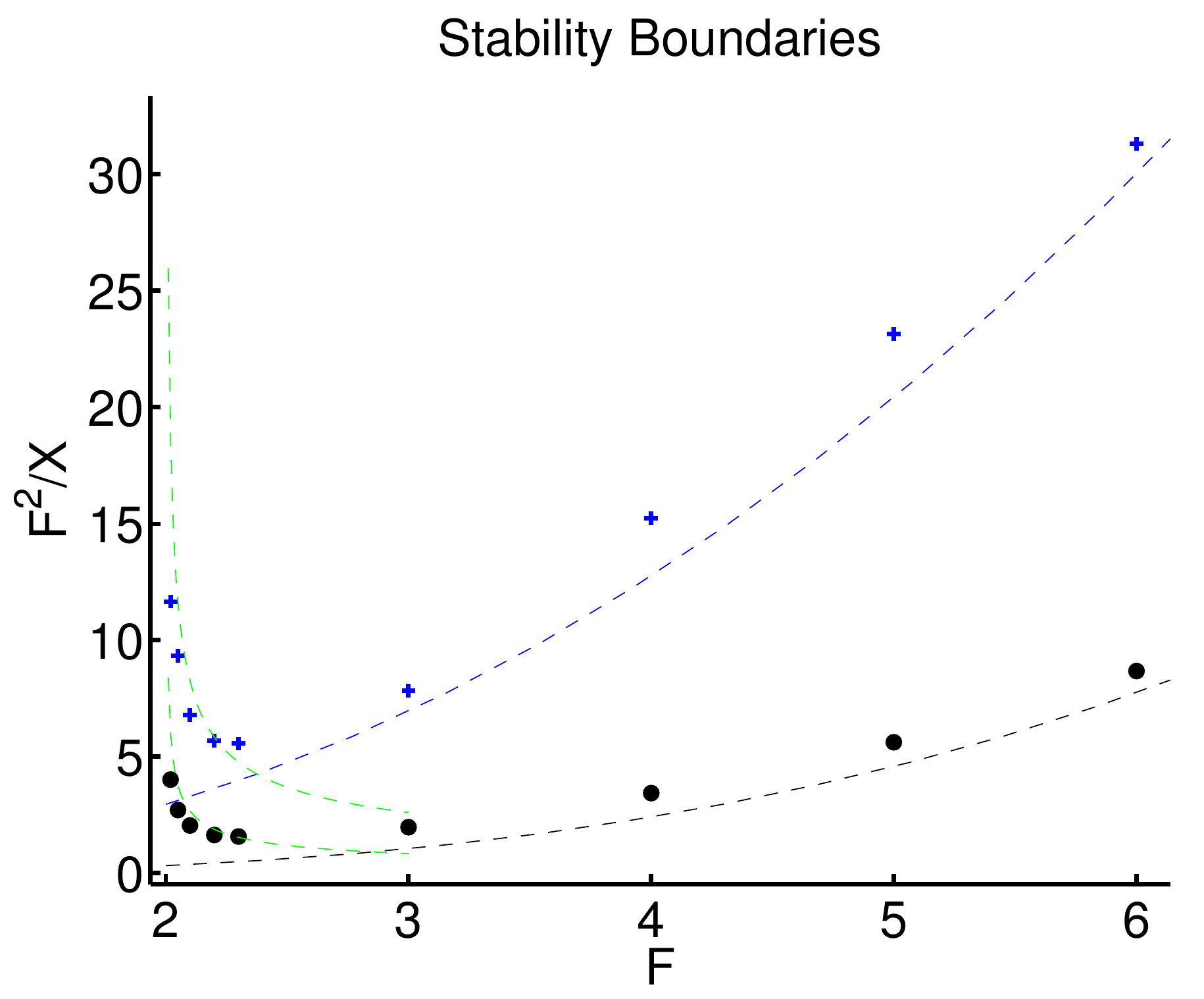}
		\end{array}
		$
	\end{center}
	\caption{ 
		Lower and upper stability boundaries for 
		$\nu=0.1$, restricted to the slice $q=0.4F$. 
		Solid dots show numerically observed boundaries. 
		Pale dashes indicate approximating curves  given by (a) (upper) $F^2/X =  e^{0.087}F^{2.88}$ and (lower) $F^2/X = e^{-2.97}F^{2.83}$, (b) (upper) $\log(F^2/X) = 2.88\log(F)+0.087$ and (lower) $\log(F^2/X) = 2.83 \log(F)-2.97$. Pale dotted curves (Green in color plates) indicate theoretical boundaries as $F\to 2^+$. (c) Small- to large-$F$ transition.}
	\label{fig476}
\end{figure}

In this section, we report on numerical investigations of (D1), (D2) and (D3) in the 
regime $2\leq F\leq 100$ (see Figure \ref{fig476})
that is relevant for hydraulic engineering applications \cite{Brock1969,Gavrilyuk2012}.
We exhibit a simple description of the stability region and find that for sufficiently large Froude numbers, stable roll-waves do not exist.
Our investigation roughly consists of two steps. 
To determine the global picture of spectrum of a linear $X$-periodic operator $L$, we use Hill's method, a Galerkin based truncation procedure which is implemented  into STABLAB \cite{Barker2009}. 
However, this method is not sufficient to study the spectrum near the origin and thus to verify hypothesis (D1), (D2) and (D3). For that purpose, we used the evaluation of the Evans function and its derivatives on contours to determine the coefficients and estimate the error terms in the 
expansion of $E_{SV}$.

A suitable parameterization, available for all Froude numbers, 
is given by $(q,X)$, where $q=-\bar c\bar\tau-\bar u$ is the {\it total outflow} and $X$ is the period. 
In \cite{Barker2015} we have gathered numerous pieces of evidence leading to the clear picture that from $F$ near $2.5-3$ and onward, stability is determined by simple relations
\[
c^-_1 \log F + c^-_2 \log q + c^-_3 \log X +c_4^- \log \nu \geq d^-
\ \textrm{and}\ 
c^+_1 \log F + c^+_2 \log q + c^+_3 \log X + c_4^+ \log \nu \leq d^+
\]
with higher and higher accuracy as $F$ increases, for some universal 
constants $c^{\#}_{j}$ and $d^{\#}$. Constants providing the lower and upper stability boundary are given approximately by $c^{\pm}_3=1$ and
\[
c^-_1 = 0.69,~~ c^-_2 = -3.5 ,~~ c^-_4 = 0.18,~~ d^-= -0.11,
\]
and
\[
c^+_1 = 0.79,~~ c^+_2 = -1.7 ,~~ c^+_4 = 0.76,~~ d^+=2.2
\]
respectively.
In Figure \ref{fig476} we illustrate this simple rule by providing one slice of the stability diagram obtained by enforcing the arbitrary constraint $q=0.4F$.

\section{\label{sec4} Stability in the limits $F\to 2$ and $F\to\infty$}

We now consider the two asymptotic regimes $F\to 2^+$ and $F\to\infty$. We provide an analytical description of the stability region in the limit $F\to 2^+$. The limit $F\to\infty$ is studied by a combination of asymptotic expansion and numerical simulations on the limit problems; roll-waves are always unstable there.

\subsection{Stability of roll-waves at onset $F\to 2^+$}

We focus on the onset of roll-waves at $0<F-2\ll 1$. Various weakly nonlinear models have been derived depending crucially on the scaling between time and space and mildly on the precise form of the diffusion term and on whether or not a vanishing viscosity regime is under consideration. In the Korteweg-de Vries regime $(\xi,\tau)=(\delta(x-3t/2), \delta^3 t)$ and in the small amplitude limit  $h=1+\delta^2\,\tilde h(\xi,\tau)$, one obtains the generalized Kuramoto-Sivashinsky equation \cite{Yu2000} (up to additional rescaling):
\begin{equation}\label{gks}
\displaystyle
\partial_\tau\tilde h+\tilde{h}\partial_\xi\tilde h+\varepsilon\partial_{\xi}^3\tilde h+\delta\left(\partial_{\xi}^2\tilde h+\partial_\xi^4\tilde h\right)=0, \quad \varepsilon>0.
\end{equation}
\noindent
The spectral and nonlinear stability of periodic traveling waves of (\ref{gks}) in the limit $\delta\to 0$ is fully 
described in \cite{JNRZ2015} and a companion paper \cite{Barker2014}. The classification of stable periodic wave can be extended to the shallow water equations (\ref{swe1}) as follows.\\

\begin{proposition}
	For $\delta=\sqrt{F-2}$ sufficiently small, uniformly for $\delta X$ on compact sets, periodic traveling waves of (\ref{swe2}) are stable for (Lagrangian) periods $X\in\frac{\nu^{1/2}}{\tau_0^{5/4}\delta}(X_l, X_r)$ and unstable for $X\in\frac{\nu^{1/2}}{\tau_0^{5/4}\delta}[X_{min}, X_r)$ and $X\in\frac{\nu^{1/2}}{\tau_0^{5/4}\delta}(X_r, X_{max}]$ where $X_{min}\approx 6.284$, $X_l\approx 8.44$, $X_r\approx 26.1$ and $X_{max}\approx 48.3$.
\end{proposition}

\vskip 0.3cm
We do not expect other stability regimes when $\delta$ is sufficiently small. Indeed, both in the regime $\displaystyle (\xi,\tau)=(\delta^{-1}(x-3t/2),\delta^{-1}t)$ and in the regime $\displaystyle (\xi,\tau)=(x-3t/2,\delta t)$  amplitude equations have been
derived from the shallow water equations indicating that
periodic waves are always unstable \cite{Kranenburg92}, \cite{Balmforth2004}. Numerical observations support this expectation.

\subsection{Infinite-Froude number limit}

To consider now the infinite-Froude number limit $F\to\infty$, we introduce a suitable rescaling in the equations and profiles with the requirements that (i) the limiting system ($F\to\infty$) be nontrivial and (ii) the limit be a regular perturbation.  This results in a one-parameter family of rescalings indexed by $\alpha\geq -2$, given by
$\displaystyle
\tau=a F^{\alpha},\: u=bF^{-\alpha/2}, \: c=c_0F^{-1-3\alpha/2},\: X=X_0F^{-1/2-5\alpha/4}$ and $q=q_0F^{-\alpha/2}$
where $a,b:\mathbb{R}\to\mathbb{R}$ and $c_0,X_0,q_0$ are real constants. Under this rescaling, we find that $X$-periodic traveling wave solutions of (\ref{swe2}) correspond to $X_0$-periodic solutions to the rescaled profile equation
\begin{equation}\label{rprof}
\displaystyle
a''= (- a^2/c_0k_0^2 \nu) 
\big(k_0  a' F^{-3/2-3\alpha/4}(c_0^2-1/ a^3) - 1
+  a(q_0 - c_0F^{-1} a)^2 - 2c_0 k_0^2 \nu ( a')^2/ a^3\big),
\end{equation}
where $b=-q_0-c_0 F^{-1}a$. Noting that the behavior of $F^{-3/2-3\alpha/4}$ as $F\to\infty$ depends on whether 
$\alpha=-2$ or $\alpha>-2$, one obtains two classes of limiting profile equations as $F\to\infty$.  An additional rescaling $Fb=\check b$ and $F^{1/2+\alpha/4}\lambda=\Lambda$ yields the associated spectral problem 
\begin{equation}\label{e:specF->infty}
\Lambda  a -c_0k_0 a'-k_0\check b'=0;\qquad
\displaystyle\frac{\Lambda \check b-c_0k_0 \check b'- k_0 (a/\bar a^3)'}{F^{3/2+3\alpha/4}}=
-\frac{2}{F}\bar a \bar b \check b - \bar b^2 a + 
\nu k_0^2 (\check b'\bar a^2 + 2 c_0 \bar a'a/\bar a^3)',
\end{equation}
where $(a,b)$ denotes the perturbation of the underlying state $(\bar{a},\bar{b})$.   
Observe that for $\alpha>-2$ the limiting profile equations, selection principles, and spectral problems are independent of the specific value of $\alpha$. 
Noting that (\ref{e:specF->infty}) is, again by design, a \emph{regular perturbation} of the appropriate limiting spectral problem as $F\to\infty$, the following sufficient \emph{instability condition} is obtained using standard perturbation techniques.\\
\begin{proposition}
	For all $\alpha\geq -2$, the profiles of (\ref{rprof}) converging as $F\to\infty$ to solutions of the appropriate limiting profile equation, are \emph{spectrally unstable} if the appropriate limiting spectral problem about the limiting profiles admit $L^2(\mathbb{R})$-spectrum in $\Lambda$ with positive real part.
\end{proposition}

\begin{figure}
	\begin{center}
		$
		\begin{array}{lccr}
		(a) \includegraphics[scale=0.22]{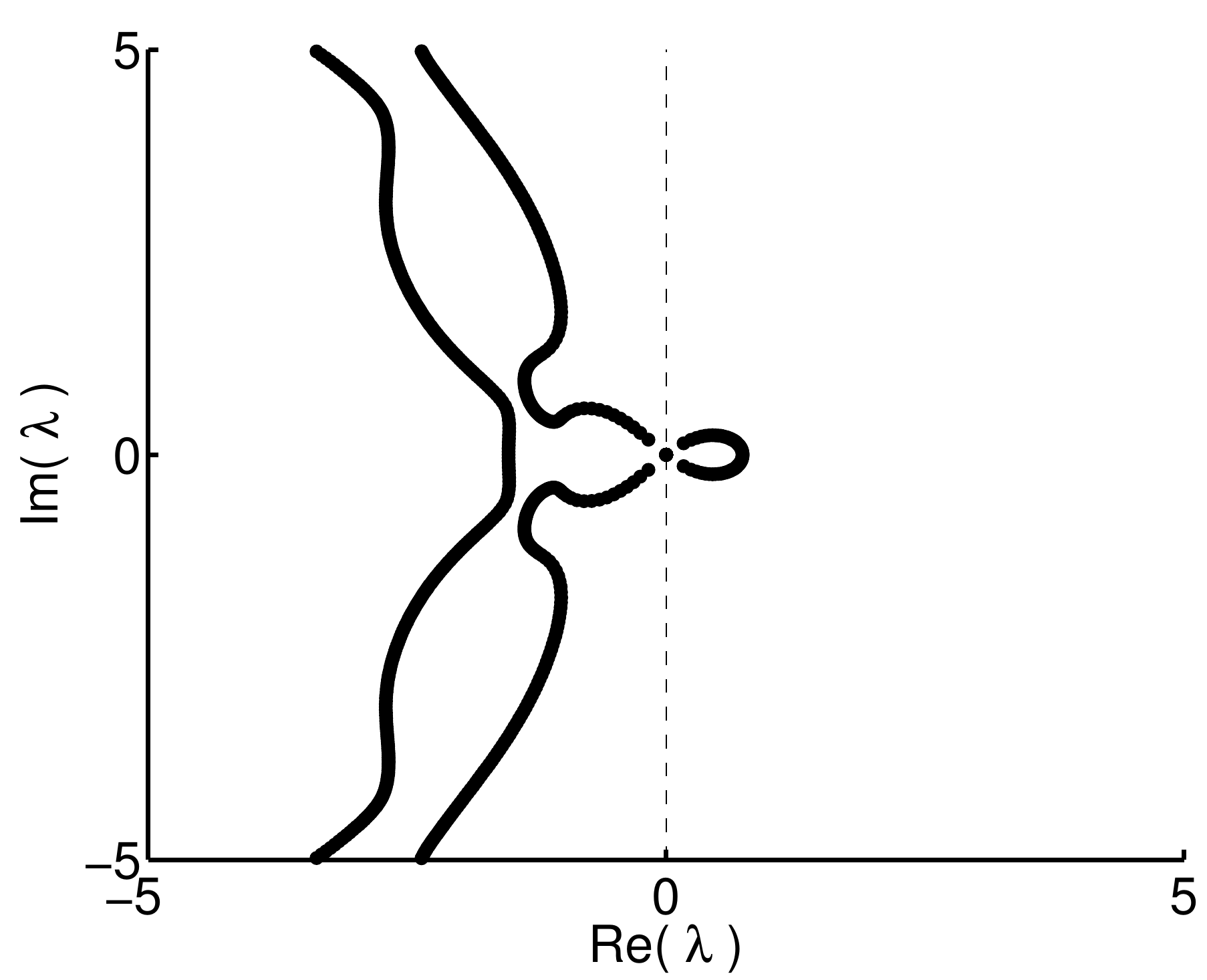}&\quad&&(b) \includegraphics[scale=0.22]{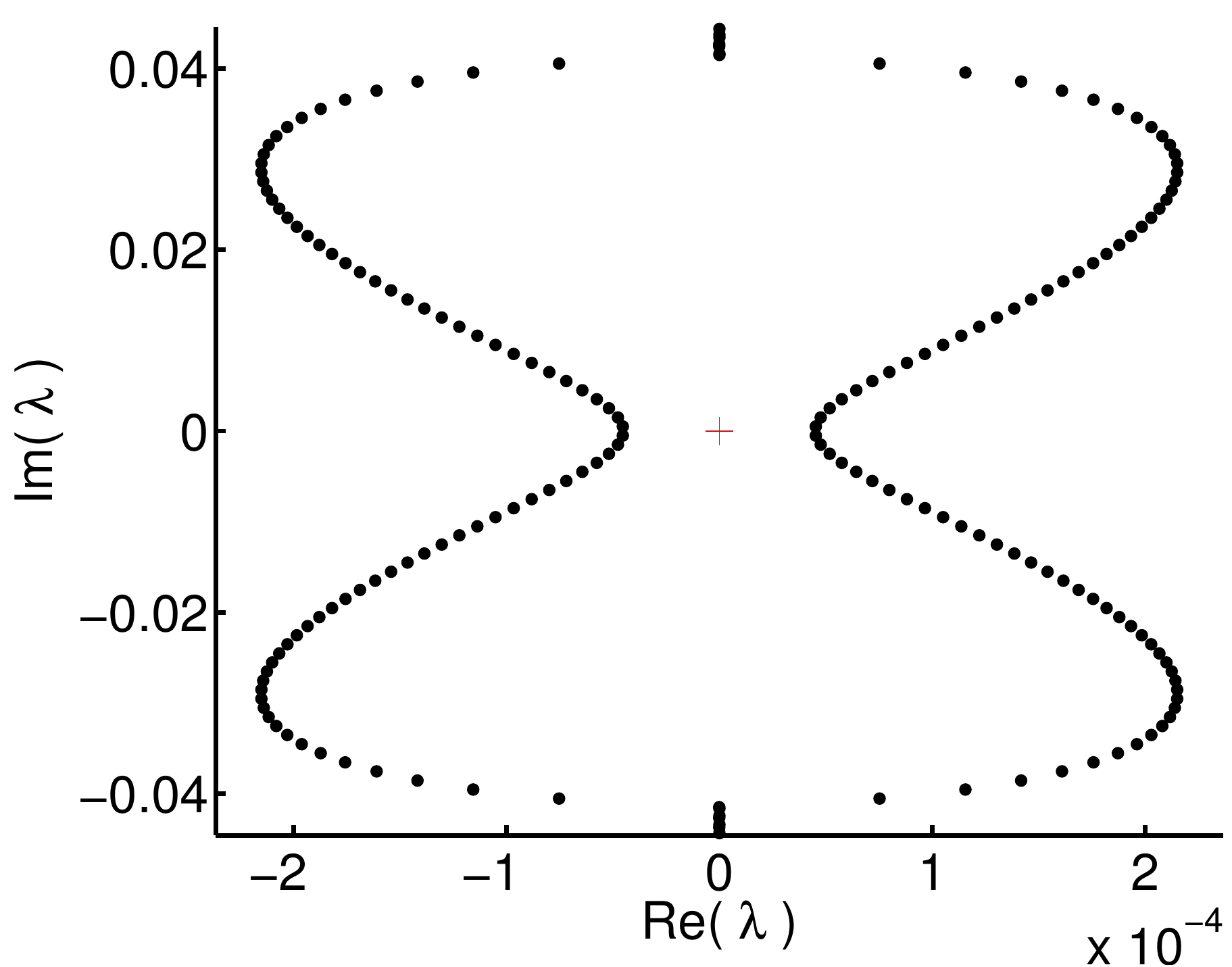}
		\end{array}
		$
	\end{center}
	\caption{In (a) and (b) we plot a numerical sampling of the (unstable) spectrum corresponding to the $F\to\infty$ limiting spectral problems
		for the cases $\alpha=-2$ and $\alpha>-2$, respectively, for a representative periodic stationary solution of the appropriate limiting profile equation.}
	\label{comparison}
\end{figure}

\vskip 0.3cm
\noindent
We have investigated the stability of the limiting spectral problems numerically in both the cases $\alpha=-2$ and $\alpha=0$; recall that the results for $\alpha=0$ in fact hold for all $\alpha>-2$.  This numerical study strongly indicates that, in both cases, all periodic solutions of the appropriate limiting profile equations are \emph{spectrally unstable} and hence spectrally stable periodic traveling wave solutions of the viscous St. Venant system (\ref{swe2}) do not exist for sufficiently large Froude numbers; see Figure \ref{comparison}.  

\medskip

\section{Conclusions and Perspectives}\label{conclusions}

We have provided a complete stability diagram in the plane $(F,q,X,\nu)$, $F$ being the Froude number, $q$ the total discharge rate, $X$ the period
and $\nu$ the Reynolds number. 
For various parametrizations of the problem, we found that for each $F\in[0, F^*)$ (for some $F^*<\infty$), $\nu>0$ and $q$ fixed in some $(F,\nu)$-dependent interval, there exist $X_{min}(F,q,\nu)$ and $X_{max}(F,q,\nu)$ such that $X$-periodic roll-waves are stable if $X\in (X_{min}(F,q,\nu), X_{max}(F,q,\nu))$. Generically, the transition to instability for $X\approx X_{min}(F,q,\nu)$ is due to a loss of hyperbolicity of the Whitham modulation equations. On the other hand, the transition to instability for $X\approx X_{max}(F,q,\nu)$ is due to the crossing of a pair of eigenvalues far from the origin and is thus undetectable by similar criteria.	

Up to now, we have considered only viscous shallow water equations with turbulent friction terms. It is an interesting and physically relevant problem to extend our results to more realistic turbulent shallow water models such as (a viscous version of) the one derived in \cite{Gavrilyuk2012} which accurately reproduces Brock's experiments on turbulent roll-waves \cite{Brock1969}. Another physically relevant problem is to consider laminar roll-waves as found e.g. in \cite{Liu1995}. In this case, we would have to take into account surface tension 
effects as they play there an important role.



\end{document}